# The *IRAS* 1.2 Jy Survey: Redshift Data[1]

Karl B. Fisher[2], John P. Huchra[3], Michael A. Strauss[2],
Marc Davis[4], Amos Yahil[5], & David Schlegel[6]

[2]Institute for Advanced Study, School of Natural Sciences, Princeton, NJ 08540
[3]Center for Astrophysics, 60 Garden St., Cambridge, MA 02138
[4]Physics and Astronomy Departments, University of California, Berkeley, CA 94720
[5]Astronomy Program, State University of New York, Stony Brook, NY 11794-2100
[6]Astronomy Department, University of California, Berkeley, CA 94720

## Abstract

We present the redshift data for a survey of galaxies selected from the data base of the *Infrared Astronomical Satellite* (*IRAS*). This survey extends the 1.936 Jy sample of Strauss *et al.* (1992a) from a flux limit of 1.936 Jy at 60 $\mu$m to 1.2 Jy. The survey extension consists of 3920 sources in the flux interval 1.2 − 1.936 Jy, of which 2663 are galaxies with measured redshifts. Fourteen objects (0.52%) do not have redshifts. The survey covers 87.6% of the sky. The data for the complete 1.2 Jy survey (the data presented here in addition to that of Strauss *et al.* 1992a) may be obtained in a machine-readable form from the National Space Science Data Center.

---

# 1 Introduction

We have completed a redshift survey of galaxies extracted from the data base of the *Infrared Astronomical Satellite* (*IRAS*)[7] Point Source Catalog (1988; PSC). The survey contains 5321 galaxies and is complete to a flux limit of 1.2 Jy at 60 $\mu$m. It represents the extension of the redshift survey of Strauss *et al.* (1992a; hereafter S92) which was flux-limited at 1.936 Jy and contained 2658 galaxies; this paper presents the subset of the redshift data in the flux interval 1.2 Jy $< f_{60} <$ 1.936 Jy ($f_\lambda$ denotes the flux density at wavelength $\lambda$) which includes 2663 galaxies.

The *IRAS* mission provided a near full sky survey based on a uniform flux scale at wavelengths unaffected by Galactic extinction; this makes the *IRAS* data base an ideal resource for constructing galaxy surveys which are free from many of the systematic biases that plague optically selected samples. The selection criteria of the 1.2 Jy survey are, with the obvious exception of the flux limit, identical to that of the 1.936 Jy sample as presented in Strauss *et al.* (1990; hereafter S90) and S92. Briefly they are:

- $f_{60} >$ 1.2 Jy, the flux limit of the survey determined using the flux densities listed in the PSC corrected by the ADDSCANing procedure explained in S90.

- $f_{60}^2 > f_{12} f_{25}$, a color restriction designed to reject many of the hotter Galactic objects (cf., S90).

- Outside excluded zones (12.34% of the sky) defined as:
  1. The Galactic plane region, $|b| < 5°$ (8.7%);
  2. Areas lacking sky coverage in PSC ($\sim$ 4%);
  3. Confused regions in PSC (0.7%).

- Moderate or high flux quality at 60 $\mu$m as listed in the PSC.

---

[7]The Infrared Astronomical Satellite was developed and operated by the U.S. National Aeronautics and Space Administration (NASA), the Netherlands Agency for Aerospace Programs (NIVR), and the U.K. Science and Engineering Research Council (SERC).

There have been a number of papers analyzing the 1.2 Jy *IRAS* survey. Fisher *et al.* (1992) examined the sample for evidence of density evolution. The acceleration of the Local Group was investigated by Strauss *et al.* (1992*b*). The power spectrum and two point correlation function in redshift space were the subject of Fisher *et al.* (1993, 1994*a*). Counts in cells statistics were examined by Bouchet *et al.* (1993). The effects of redshift distortions on clustering were analyzed in Fisher *et al.* (1994*b*), Fisher, Scharf, & Lahav (1994), Cole, Fisher, & Weinberg (1994, 1995), Fisher *et al.* (1995), Nusser & Davis (1994), and Heavens & Taylor (1995).

## 2 Observations

The spectroscopic observations of the objects in our candidate list with $1.2 < f_{60} < 1.936$ Jy were carried out between July, 1989 and January, 1991 on five different telescopes. The brightest objects in the Northern sky ($\delta > -20°$) were observed by J.P.H. with the 1.5-m Tillinghast reflector of the Whipple Observatory on Mount Hopkins; some fainter objects were observed by J.P.H. on the Multiple Mirror Telescope. The remainder of the Northern objects were observed by K.B.F, M.D., and M.A.S. on the Lick Observatory 3-m reflector. The majority of objects in the Southern sky were observed by K.B.F., M.D., and M.A.S. on the 1.5-m telescope at the Cerro Tololo Inter-American Observatory (CTIO); we also had one three night run at CTIO on the 4-m telescope. In order to minimize the number of spectroscopic observations, we periodically compared our candidate list with those in the literature (Huchra *et al.* 1992, hereafter ZCAT); this resulted in a total of 923 matches with 1.2 Jy $< f_{60} < 1.936$ Jy.

Observations at Lick Observatory were performed using the 3-m Shane telescope with a 800 × 800 pixel TI CCD Cassegrain spectrograph (Lauer *et al.* 1983). The optical configuration on the 3-m consisted of a transmission grating blazed at 6500Å with 600 lines mm$^{-1}$. In our observations, we used a long slit of width 2″ usually oriented in the E-W direction; occasionally, however, we rotated the position angle of the slit to allow multiple objects to be observed simultaneously. The resolution of the setup was 5Å. Spectroscopy was performed in the red with a typical wavelength coverage of 5800−8000Å. Most of the *IRAS* galaxies in our sample have strong line emission at H$\alpha$ $\lambda$6563, [N II]$\lambda\lambda$6548,6583, and [S II]$\lambda\lambda$6716,6731, so this choice of wavelength coverage allowed detections of redshifts $z \lesssim 0.2$. Roughly 30 high redshift objects were detected from line emission in [O III]$\lambda\lambda$4959,5007 and from absorption in the NaD $\lambda$5876 line; our spectral coverage was always adequate to ensure that at at least one of these lines was available for any redshift. Comparison spectra were taken after each observation using He-Ne-Ar lamps. Typical exposure times ranged from 5 to 15 minutes, depending on the weather and the optical brightness of the object. Seeing at Lick was $\sim 1.5 - 2''$.

The instrumentation for the CTIO 1.5-m telescope consisted of a 576 × 421 pixel GEC CCD Cassegrain spectrograph and a reflection diffraction grating blazed at 6750Å with 300 lines mm$^{-1}$. Again, we used a long slit of width 2″ oriented in the E-W direction. Exposure times varied from 5 to 25 minutes. The resolution was 8Å. A similar setup was used during the single run on the CTIO 4-m telescope and consisted of a 576 × 421 pixel GEC CCD spectrograph with a reflection grating blazed at 7500Å with 316 lines mm$^{-1}$. The slit size and orientation on the 4-m were the same as on the 1.5-m. Resolution on the 4-m was 6Å. As at Lick, spectra were taken in the red ($\lambda \sim$ 5800–7800Å) and comparison lamps were taken after each exposure. The objects observed on the 4-m run were the faintest optical galaxies in the survey and exposure times were typically 20 minutes. Seeing at CTIO was somewhat better than at Lick, typically $\sim 1 - 1''.5$.

At Whipple Observatory, data were taken on the 1.5-m using a Reticon photon-counting tube (Latham 1982) with a 3″.2 × 6″.4 aperture and a resolution of 6Å. The setup on the MMT also used a photon-counting Reticon but was preceded by an S20 EMI image tube. A grating with 300 lines mm$^{-1}$ was used, giving a spectral coverage from 3200-7200Å and a resolution of 8Å.

## 3  Data Reduction

The data reduction was performed using a procedure outlined in S92; we give a brief summary here for completeness.

We began each night of observing at Lick and CTIO by obtaining a series of dome flats. This procedure consisted in taking a series of short exposures of a white spot inside in the dome which was illuminated by a quartz lamp. We removed large scale gradients in our flat field by compressing the median flat into a single column and performing a spline fit to the resultant averaged column. We then divided the median flat, column by column, by this spline fit. The resulting image became our final flat field, free of any large scale gradients, and with a typical pixel to pixel variation of 10%.

The two-dimensional spectra were then bias subtracted and flat fielded in the usual way. Cosmic rays were removed in the two-dimensional images by interpolation using the Berkeley version of the program VISTA (Ebneter 1989). One-dimensional spectra were extracted and sky subtraction performed, again using VISTA. On one run on the CTIO 1.5-m, we used the IRAF[8] software package to perform these procedures.

We fitted a polynomial to the centroids of the emission lines in the comparison lamp spectra, typically of fifth order. There were ~10 lines used in the fits whose residuals had an $rms$ dispersion of $\lesssim 0.3$Å. This fit was then used to wavelength calibrate the galaxy spectrum taken immediately prior to the comparison lamp spectra. Gaussian line profiles were simultaneously fit to H$\alpha$ and the [NII] and [SII] doublet emission lines in the wavelength calibrated spectrum. In a few cases, Gaussian profiles were fit to the NaD absorption line or the [OIII] emission line.

Redshifts obtained from the fitted lines were then corrected to the heliocentric frame. Errors were estimated from the uncertainties in the fitting procedure, residuals in the fits to the comparison lamp spectra, and a slit guiding term. The slit guiding term was computed assuming a point source viewed through the slit with typical seeing. The resulting slit guiding error was 22 km s$^{-1}$ for the Lick 3-m telescope and 35 km s$^{-1}$ for the CTIO 1.5-m (both computed assuming 2″ seeing). For the one run on the CTIO 4-m, we assumed 1.5″ seeing and derived a guiding error of 39 km s$^{-1}$.

Data reductions for observations obtained by J.P.H. on the Tillinghast and MMT telescopes were similar to that described above. In addition to fitting to emission lines,

---

[8]IRAF is distributed by the National Optical Astronomy Observatories, which is operated by the Association of Universities for Research in Astronomy, Inc. (AURA) under cooperative agreement with the National Science Foundation.

J.P.H. obtained some redshifts by cross-correlating absorption lines with stellar velocity standards using the method of Tonry & Davis (1979).

The median redshift error for galaxies observed at Lick and CTIO is 38 km s$^{-1}$, with 90% having errors < 75 km s$^{-1}$. We did not try to ascertain the reliability of redshifts taken from the literature and simply adopted the published error when available.

## 4  Results of Observations and Data Reduction

Observations for the 1.2 Jy catalog were completed in January, 1991. Of the 8934 sources in the PSC that satisfy our selection criteria, 5321 are galaxies for which we obtained redshifts; the data of the 5014 sources (and redshifts of 2658 galaxies) with $f_{60} \geq 1.936$ Jy are given in S92. Currently, 14 objects (0.26% of the galaxy sample) in the 1.2 Jy sample remain unobserved; most of these are galaxies from their appearance on the sky survey plates. The positions and fluxes of the unobserved objects are listed in Table 2. The classifications of the objects in our PSC candidate list are given in Table 1. We describe how objects were classified as galaxies and non-galaxies in § 4.2. As we discuss there, the classification of Galactic objects inTable 1 is in no way complete, with the vast majority of classifications based on their visual appearance on sky survey plates rather than spectra.

The upper plot of Figure 1 shows the sky distribution, in an Aitoff projection in Galactic coordinates, of the 5321 galaxies in the sample. The distribution of Galactic sources in the catalog is shown in the lower plot of Figure 1. This figure should be compared with the similar plots of S90 (Figures 16 and 17, although note that the handedness of these plots is the opposite of those). The Galactic objects are concentrated near the Galactic Plane with particularly heavy concentrations near the Orion-Taurus ($l = 210°$, $b = -20°$) and Ophiuchus ($l = 345°$, $b = 20°$) complexes.

### 4.1  Color Corrections

The detectors aboard *IRAS* measured the flux, $F$, for each object in four broad bandpasses. The PSC actually quotes flux "densities," calculated from $F$ assuming that the spectral energy distribution (SED), $f_\nu$, is proportional to $\nu^{-1}$, i.e.,

$$F = f_\nu^{PSC}(\nu_\circ) \int (\nu_\circ/\nu) R_\nu d\nu \quad , \qquad (1)$$

where $\nu_\circ$ is the frequency at the bandpass center and $R_\nu$ is the responsivity of the detector. Consequently, if the SED is not inversely proportional to $\nu$, then the quoted PSC flux density is not an accurate estimate of the source's actual flux density.

Fisher *et al.* (1992) discuss how to correct the quoted PSC flux densities given a model for the SED. Unfortunately, since *IRAS* only provides estimates of the SED at four frequencies, the modeled SED, and hence the color correction, is inherently uncertain. Moreover, since we do not apply any flux quality restrictions at 12, 25, and 100 $\mu$m, there is no guarantee that the fluxes at these wavelengths will be actual detections instead of upper limits. Indeed, most galaxies in our sample are undetected by *IRAS* at 12 or 25 $\mu$m. Fisher *et al.* (1992) used two different fits to the SED, one based on a gray-body fit to the fluxes at 60 and 100 $\mu$m, and one based on a polynomial fit to the fluxes at 25, 60, and 100 $\mu$m. Figure 2 is a histogram of the fractional change in the 60 $\mu$m fluxes determined using the polynomial model for the SED; the corrections in this case are small, $\lesssim 4\%$.

We have looked for correlations between the magnitude of individual color corrections with both local density and sky position. No correlations were found, thus suggesting that the bias in large-scale structure studies introduced by neglecting the color corrections is, if present, very small. There is a weak effect with redshift, as there is a correlation between color and luminosity; however, the effect of this is small (Fisher *et al.* 1992). In our analyses of the sample, unless otherwise stated, no color corrections to the fluxes have been performed.

Infrared Galactic cirrus at low Galactic latitudes is our principal source of contamination. It tends to be cooler than external galaxies, in the sense of having a higher ratio of 100 to 60 micron flux. This difference in spectral colors has been used by some groups to help exclude cirrus from galaxy candidate lists drawn from the PSC (e.g., Rowan-Robinson *et al.* 1991). However, one cannot make a clean distinction between external galaxies and Galactic cirrus based on *IRAS* colors alone (Figure 3), and we decided not to adopt a color restriction on the 100 to 60 $\mu$m flux ratio for the 1.2 Jy survey. Figure 3 shows the distribution of $f_{100}/f_{60}$ flux ratios for galaxies and cirrus sources in the 1.2 Jy sample based on optical identifications from sky survey plates and spectroscopic observations (§ 4.2).

## 4.2 Completeness

As discussed in S90, our color criterion, $f_{60}^2 > f_{12} f_{25}$, is very good at rejecting stars but unfortunately fails to reject many Galactic cirrus sources. 3920 objects passed our color criteria and had 1.2 Jy $< f_{60} <$ 1.936 Jy; we expected about 2500 of these sources to be actual galaxies with the rest being mostly cirrus. In order to pare down the list of objects needing spectroscopic observations, we wanted a way to reject many of the cirrus sources prior to going to the telescope.

Meiksin & Davis (1986; hereafter MD) constructed an *IRAS* galaxy sample using much more restrictive selection criteria than we have adopted for the 1.2 Jy sample. They insisted that galaxy candidates be non-variable, have a highest quality flux flag at 60 $\mu$m, have $f_{60}/f_{12} > 3$, and not have a catalog match in the PSC with a Galactic object. Furthermore, they defined a series of excluded regions around the Magellanic Clouds, and regions of very high source density at low Galactic latitudes.

Strauss (1989) found that the MD selection criteria had an efficacy of greater than 99% for selecting galaxies. S90 examined by eye all objects in the candidate list which were in the MD excluded zones on the ESO and POSS Sky Survey Plates with a magnifier, and removed those objects which were not resolved. The same procedure was followed for the extension to 1.2 Jy. We erred on the side of caution, including for telescopic follow-up any object that was not unambiguously unresolved.

This procedure, while necessary to remove objects clearly associated with Galactic sources, will lead to some incompleteness. Sources in regions of high Galactic extinction may be invisible altogether on the Sky Survey prints. In addition, many starburst galaxies in the sample are quite nucleated; they could be indistinguishable from stars if a moderate amount of Galactic extinction made their faint disks invisible. It also may be possible to improve our "by-eye" examination of the ESO and POSS plates by obtaining more accurate source identifications, e.g., by using optimal identification methods based on likelihood ratios (Sutherland & Saunders 1992).

The objects that were not rejected from examination of the plates (including *all* objects outside the MD excluded zones) were taken to the telescope. We obtained CCD images of many of the remaining ambiguous objects in $R$ band using the 40" Nickel telescope at Lick Observatory, and the 0.91-m telescope at CTIO; this allowed us refine the observing lists further. For those ambiguous objects for which we did not have CCD images, inspection of the fields on the acquisition camera of the spectrograph was usually enough to identify the galaxy candidate, or to confirm that none existed. Finally, in remaining ambiguous fields, we took spectra of several objects in the *IRAS* error circle to identify the *IRAS* source, be it Galactic or extragalactic. We emphasize that essentially all these ambiguous fields lay close to the Galactic plane.

One possible completeness check is to compare the sky distribution of observed galaxies in the sample with a corresponding estimate of the background cirrus intensity. If *the galaxy and cirrus distributions are independent*, then one can obtain an estimate of completeness by looking at the distribution of galaxy counts as a function of cirrus intensity, looking for the completeness to drop in regions of highest cirrus intensity. A convenient estimate of the background cirrus intensity is given in the PSC in the form of a flag, CIRR3 (*IRAS* Explanatory Supplement 1988, § V.H.4; hereafter ES). We made a map of the cirrus emission by averaging the CIRR3 flags for all objects in the PSC on an equal area Aitoff projection in Galactic coordinates (ES § X.D.2.b). We then looked for a deficit of galaxy counts in regions of high cirrus intensity. Assuming that the galaxy counts at high Galactic latitudes, $|b| > 30°$, were complete, we estimated the completeness of the survey at $|b| < 10°$ to be $\sim 97\%$, that is, about 20 objects missing from the sample. However, the assumption that cirrus density is uncorrelated with galaxy density is not valid in detail; there is an enhancement of nearby large scale structure (e.g., the "Great Attractor" and Perseus-Pisces superclusters) at low Galactic latitudes (cf., Figure 6).

# 5 Sample Overview

## 5.1 Redshift Distribution and Selection Function

In any flux-limited catalog of galaxies, the number density of objects decreases with distance. This is simply because at large distances fewer objects have luminosities sufficiently large that their flux falls above the flux limit of the catalog. The mean number density of galaxies in a homogeneous universe lying in a shell at distance $r$ of width $\Delta r$ is given by $\bar{n}\omega r^2 \phi(r)\Delta r$, where $\phi(r)$ is the *selection* function of the sample, $\omega$ is the solid angle subtended by the sample, and $\bar{n}$ is the mean density of the sample (cf., Equation 4 below). Explicitly, $\phi(r)$ is given by an integral of the luminosity function, $\Phi(L)$, over all luminosities bright enough to lead to fluxes above the flux limit, $f_{min}$,

$$\phi(r) = \frac{\int_{4\pi r^2 \nu f_{min}}^{\infty} \Phi(L) dL}{\int_{4\pi r_s^2 \nu f_{min}}^{\infty} \Phi(L) dL} \quad . \tag{2}$$

where $r_s = 500$ km s$^{-1}$ is a fiducial redshift below which the selection function is set to unity.

Yahil *et al.* (1991) show that a two power law functional form is both a robust and accurate parameterization of the *IRAS* galaxy selection function:

$$\phi(r) = \begin{cases} 1 & \text{if } r < r_s; \\ \left(\frac{r_s}{r}\right)^{2\alpha} \left(\frac{r_*^2 + r_s^2}{r_*^2 + r^2}\right)^{\beta} & \text{if } r \geq r_s. \end{cases} \tag{3}$$

The parameters of the selection function $\{\alpha, \beta, r_*\}$ in Equation 3 are determined using a maximum likelihood estimator introduced by Sandage, Tammann, & Yahil (1979) which is unbiased in the presence of density inhomogeneities. For the 1.2 Jy survey with redshifts measured in the Local Group (LG) frame (determined using the transformation of Yahil, Tammann, & Sandage 1977), the best fit parameters determined from the galaxies within 10,000 km s$^{-1}$ are:

$$\alpha = 0.483 \pm 0.070 \quad \beta = 1.790 \pm 0.134 \quad r_* = 5034 \pm 723 \text{ kms}^{-1} \tag{4}$$

These parameters have a normalized covariance matrix given by:

$$\begin{array}{c|ccc} & \alpha & \beta & r_* \\ \alpha & 1.00 & 0.09 & 0.76 \\ \beta & 0.09 & 1.00 & 0.64 \\ r_* & 0.76 & 0.64 & 1.00 \end{array} \tag{5}$$

Given a determination of the selection function, one can determine the mean density of the sample:

$$\bar{n} = \frac{1}{V} \sum_i \frac{1}{\phi(r_i)} \quad , \qquad (6)$$

where $V$ is the volume of the survey to 10,000 km s$^{-1}$ and the sum is over the galaxies contained in the sample. We find $\bar{n} = 0.062 \pm 0.0016 \ h^{-3} \mathrm{Mpc}^{-3}$, corresponding to the mean density of galaxies bright enough to pass our flux limit at $r = r_s$, i.e., the density of galaxies with luminosities $L \geq 4\pi r_s^2 \nu f_{min}$. The luminosity function, $\Phi(L)$, can be written in terms of the derivative of the selection function (cf., Equation 2); the resulting luminosity function is qualitatively very similar to the luminosity function of the 1.936 Jy sample as shown in Figure 7 of Yahil *et al.* (1991).

The redshift distribution of the survey is shown in Figure 4. The histogram is in linear bins of 200 km s$^{-1}$ based on LG recession velocities. The solid line is the predicted counts in a homogeneous universe characterized by the selection function in Equation 3 with parameters given in Equation 4. The median LG redshift of the sample is 5859 km s$^{-1}$.

## 5.2 Sky Distribution

Figures 5–8 show the distribution of galaxies in the 1.2 Jy survey on the sky in Galactic coordinates. Each figure shows a different redshift interval (in the LG frame). Each figure has two panels. The upper panel shows the actual galaxy distribution within a specified redshift interval in an equal area Aitoff projection in Galactic coordinates. The lower panel shows the smoothed galaxy distribution on a spherical shell through the center of the redshift interval. The smoothing is performed in three dimensions with a Gaussian whose dispersion is the local mean galaxy separation, i.e., $l_S = 1/[\bar{n}\phi(r)]^{1/3}$ (determined from the selection function and mean density in Equations 3 and 4); the full width at half maximum of the Gaussian smoothing window ($2.35 \, l_S$) in each case is represented by the solid bar. The excluded regions were interpolated using the methods of Yahil *et al.* (1991) to make the contour plots.

The upper panel of Figure 5 shows all galaxies with LG redshifts in the interval $0 < v < 2500$ km s$^{-1}$ The lower panel shows the smoothed galaxy distribution on a shell at redshift of 1250 km s$^{-1}$. Notice that the nearby galaxy distribution is quite anisotropic on the sky. The dominant feature in this interval is the Supergalactic Plane (SGP) defined by de Vaucouleurs, de Vaucouleurs, & Corwin (1976), which in this projection appears as an overdensity along $l \approx 140°$ and $l \approx 300°$ in the Northern hemisphere. The Virgo supercluster is the prominent overdensity at ($l = 284°$, $b = +75°$), while the Ursa Major cloud stands out at ($l = 145°$, $b = +65°$). The Local Void (Tully & Fisher 1987) is apparent as the deficit of galaxies in the longitude interval $l = 0°$ to $l = 120°$. The Fornax and Eridanus clusters are the overdensity at $170 < l < 240°$, $-60 < b < -45°$.

The distribution of galaxies within the interval 2500 km s$^{-1} < v <$ 5000 km s$^{-1}$ is shown in Figure 6. The so-called "Great Attractor" region stands out as the structure at ($l = 300 - 360°$, $b = -45 - +45°$) (cf. Burstein, Faber, & Dressler 1990). This region is composed of the Hydra-Centaurus supercluster complex at ($l = 300 - 360°$, $b = 0 - +45°$) and the Pavo-Indus supercluster at ($l = 320-360°$, $b = -45-0°$). The extended overdensity at this redshift is the portion of the Perseus-Pisces supercluster ($l = 120 - 160°$, $b = -30 - +30°$). The overdensity at ($l = 210°$, $b = -30°$) is the NGC 1600 group; this group is also prominent in the Optical Redshift Survey of Santiago *et al.* (1995).

Figure 7 shows the galaxy distribution in the range 5000 km s$^{-1} < v <$ 7500 km s$^{-1}$. The SGP no longer stands out. The only noticeable overdensity is that associated with the high-redshift component of the Perseus-Pisces supercluster. In Figure 8, we plot galaxies in the interval 7500 km s$^{-1} < v <$ 10000 km s$^{-1}$. At this distance the sky looks fairly homogeneous, although the sampling is becoming much more dilute, and the smoothing length is much larger.

## 5.3 The Local Density Field

The density field of *IRAS* galaxies has been previously presented in several papers. Strauss *et al.* (1992c) show contour maps of the density field of the original 1.936 Jy sample. Saunders *et al.* (1991) present a beautiful series of color images of the density field derived from a redshift survey of one in six *IRAS* galaxies to a flux limit of 0.6 Jy (QDOT), and Strauss & Willick (1995) present the density field of the 1.2 Jy survey. The 1.2 Jy survey has a greater sampling density than the QDOT survey out to $\sim$ 20,000 km s$^{-1}$, allowing higher resolution. We present the density field here in a sequence of slices through the smoothed density field. As in the lower panels in Figures 5–8, the smoothed density field was derived by convolving the galaxy redshift distribution with a Gaussian with a dispersion given by the mean galaxy separation. The use of a variable smoothing length maintains high resolution in the well-sampled nearby density field, while preventing undersampling at large distances, although it means that the contrast in the maps is a decreasing function of distance from the origin. The density field presented here is in *redshift* space; only a transformation to the Local Group rest frame has been applied.

Figure 9 shows the smoothed density fluctuation field, $\delta\rho/\rho$, in a series of slices parallel to the SGP (Z=0), spaced by 2000 km s$^{-1}$. The SGP is the middle panel (e). The contours correspond to $\Delta\delta\rho/\rho = 0.25$, with dashed contours corresponding to negative values. The mean density contour is heavier than the others. The planes in the figure correspond to offsets relative to the SGP of a) Z = $-8000$, b) $-6000$, c) $-4000$, ..., g) 4000, h) 6000, and i) 8000 km s$^{-1}$.

The Virgo supercluster and the Ursa Major cloud appear as a single prominent elongated overdensity in the center of the middle panel of Figure 9. In this plane, the Hydra-Centaurus (H-C) supercluster complex (the "Great Attractor") is the large overdensity at X = $-3500$ km s$^{-1}$ and Y = 1000 km s$^{-1}$, while the Perseus-Pisces (P-P) supercluster is stands out on the region of the sky opposite of H-C at (X = 4500 km s$^{-1}$, Y = $-2000$ km s$^{-1}$). Both the P-P and H-C superclusters are sandwiched between large voids. The small overdensity at (X$\sim$ 0 km s$^{-1}$, Y$\sim$ 6000 km s$^{-1}$) is A1367.

The plane to the left of the central panel in Figure 9 is located at Z = $-2000$ km s$^{-1}$. In this plane, the P-P supercluster is dominant while only the periphery of the Hydra supercluster remains visible. In the plane at Z=2000 km s$^{-1}$ the single large overdensity is the Pavo-Indus-Telescopium supercluster complex, contiguous with H-C. The Coma cluster is visible in this plane as the weak overdensity at (X $\sim$ 1000 km s$^{-1}$, Y $\sim$ 8000 km s$^{-1}$). The Local Void of Tully & Fisher (1987) is particularly striking in this plane; it is seen to be contiguous with a void extending to the edge of the survey volume at this value of Z.

The "Great Wall" of the CfA survey (Geller & Huchra 1989) can be seen in panels e), f), and g) of Figure 9 as the overdensity at X$\sim$ 0 km s$^{-1}$ and Y$\sim$ 7500 km s$^{-1}$.

# 6 Data Presentation

The galaxy data with $1.2 < f_{60} < 1.936$ Jy are presented in Table 3; it contains 2663 objects with redshifts. The columns listed are as follows:

- *Columns 1, 2:* Right Ascension and Declination of the galaxy, epoch 1950. These positions are taken from the *IRAS* PSC and from Rice *et al.* (1988) for resolved galaxies which are not listed in the PSC.

- *Column 3:* The 60 $\mu$m flux of the source in Jy. Fluxes followed by the letter A are estimated by obtaining ADDSCAN's of the *IRAS* position from the raw *IRAS* data base; these sources are those flagged as extended, variable, or of low flux quality (cf., S90). Otherwise, fluxes are drawn from the PSC.

- *Column 4:* The optical identification of the source if it exists in the standard optical galaxy catalogs. These listings are not necessarily complete; the absence of a name in this column does not necessarily imply that the source does not exist in the optical catalogs. For those sources that are found in more than one galaxy catalog, the name is chosen from that catalog highest in the following list: NGC or IC, Zwicky, UGC (Nilson 1973) or ESO (Lauberts 1982) (zeroes are added to the names of the latter to fill in blanks). Objects not appearing in one of these catalogs are given names from the MCG and other catalogs.

- *Column 5:* The redshift of the source as measured in the heliocentric frame. The units are km s$^{-1}$. Redshifts are defined in the optical convention; $cz = 299,792$ km s$^{-1}$ $\times$ $(\lambda - \lambda_0)/\lambda_0$. We flag sources which had more than one match within $1'.5$ in J.P.H.'s compilation of redshifts, with an 'M' preceding the redshift. In these cases, we list in this table the redshift of the galaxy that is either closest to the *IRAS* position, or which has the brightest optical magnitude.

- *Column 6:* The redshift error, as either quoted in the redshift source, or the $1 - \sigma$ error determined by us (cf., § 3). The units are km s$^{-1}$; we have not tried to assess the reliability of errors drawn from the literature. Those galaxies which we have assigned to the Local Group are flagged with an 'L' proceeding the redshift.

- *Column 7:* The reference for the redshift. These are described in Table 3. We have not endeavored to supply references from which we drew less than 10 redshifts; these are generically referenced to ZCAT. Reference lists for ZCAT galaxies can be obtained via anonymous ftp to fang.harvard.edu in /pub/catalogs/zsource.tex.

Table 4 contains data on the non-galaxies included in the sample with $1.2 < f_{60} < 1.936$ Jy, and contains 1237 objects. The sky distribution of these objects is given in Figure 1. Although some objects have been observed at the telescope, the majority have been rejected based on their appearance on the Sky Survey plates and in a few cases from imaged fields (cf., § 4.2). Thus our classifications (cf., Table 1) for these objects are very incomplete. The table consists of the following:

- *Columns 1,2:* Right Ascension and Declination from the PSC.

- *Column 3:* Flux density in Jy at $60\,\mu$m in Jy. Fluxes derived from ADDSCANs are marked with the letter A; unlike the galaxies in Table 3, the ADDSCANing for extended, variable, and moderate flux quality sources is not complete for the non-galaxies in this table.

- *Column 4:* The nature of the source. This is coded as a single letter, as follows:

C "Cirrus" source. Most of these objects were not observed at the telescope, but were rejected based on their appearance on the Sky Survey prints.

H Resolved HII region in our own or a nearby galaxy. The flux from the galaxy as a whole is taken from Rice *et al.* (1988) and thus includes the contribution from these bright objects.

P Planetary nebula. These objects have extremely strong and unmistakable emission lines. None of these planetary nebulae are new discoveries; all are listed in Perek & Kohoutek (1967) or one of the followup supplementary catalogs of planetary nebulae in the literature.

R Reflection nebula around a star.

S Star.

s Emission-line star (usually with H$\alpha$ emission). Many of these objects are accompanied by nebulae, and might be better classified as compact H II regions.

Again it must be emphasized that the classifications of these different types is very incomplete, especially at low Galactic latitudes.

A machine-readable version of the 1.2 Jy catalog can be obtained from the National Space Science Data Center (NSSDC) at the following address:

telnet nssdca.gsfc.nasa.gov (internet), or set host nssdca (decnet).

The catalog and associated files have been assigned Astrophysics Data Center (ADC) number 7185.

There are a multitude of people which deserve thanks in a project of this size. We would especially like to thank the IPAC staff and in particular Tom Soifer, Tom Chester, Elizabeth Smith, Linda Fullmer, George Helou, and Gene Kopan. At the telescopes, our work has been made easier by the able and friendly assistance of Jorge Bravo, Manuel Hernandez, Daniel Maturana, Ricardo Venegas, Steve Heathcote, Patricio Ugarte, Mauricio Fernandez, Ramón Galvez, and Hernán Tirado at Cerro Tololo, Wayne Earthman, Keith Baker, Jim Burrous, and John Morey at Lick Observatory, and Jim Peters, Ed Horine, and Janet Robertson at Mt. Hopkins. In addition to the many people cited in the references, we would like to thank H. Rood, L. da Costa, D. Latham, J. Tonry, S. Hurley, T. Fairall, J. Menzies, E. Olson, G. Bothun, and P. Chamaraux for kindly providing us with unpublished redshifts. The Time Allocation Committees of these observatories are thanked for their very generous support of this project. We thank the Astronomical Data Center at the NASA-Goddard Space Flight Center for providing us with tapes of the *IRAS* PSC Version 2. CCD and software development at Lick Observatory are supported by NSF grant # AST-86-14510. KBF acknowledges the support of a Department of Education Graduate Fellowship, a SERC Post-Doctoral Fellowship, and the Ambrose Monell Foundation. JPH acknowledges the support of NASA grant NAGW-201. MAS acknowledges the support of the WM Keck Foundation. In addition, this work has been supported under the *IRAS* extended mission program, as well as grants from the NSF and NASA.

**Author's E-mail Addresses:**

*fisher@guinness.ias.edu*

*huchra@cfa.harvard.edu*

*strauss@guinness.ias.edu*

*marc@coma.berkeley.edu*

*ayahil@sbast3.ess.sunysb.edu*

*schlegel@magicbean.berkeley.edu*

# Figure Captions

**Figure 1**: Upper panel shows the sky distribution of the 5321 galaxies in the 1.2 Jy survey in an Aitoff projection in Galactic coordinates. Lower panel shows the 3591 Galactic objects in the survey which passed our sample criteria.

**Figure 2**: Histogram showing the magnitude of the color correction for all galaxies in the 1.2 Jy survey with high quality fluxes at 25, 60, and 100 $\mu$m.

**Figure 3**: Distribution of the 100 to 60 $\mu$m flux ratio for objects in the 1.2 Jy survey. The solid curve refers to sources identified in the 1.2 Jy as galaxies, while the dashed curve represents sources classified as Galactic cirrus.

**Figure 4**: Redshift distribution of the 1.2 Jy survey. Redshifts have been corrected to the Local Group frame. The solid curve represents the expected counts in a homogeneous universe as given by the selection function.

**Figure 5**: Upper panel: Galaxy distribution in the 1.2 Jy Survey with LG recession velocities within 0 km s$^{-1}$ < $v$ < 2500 km s$^{-1}$. Lower panel: The smoothed Galaxy distribution on a shell at redshift 1250 km s$^{-1}$. Smoothing is performed in three dimensions with a variable Gaussian smoothing length given by the local mean galaxy separation. The heavy contour denotes the mean density. The dashed contours represent densities below the mean (spaced at one-third and two-thirds of the mean). Contours in excess of the mean are spaced logarithmically; each three contours corresponds to the density changing by a factor of two. The solid bar is the full width at half maximum of the Gaussian smoothing window.

**Figure 6**: Upper panel: Galaxy distribution in the 1.2 Jy Survey with recession velocities within 2500 km s$^{-1}$ < $v$ < 5000 km s$^{-1}$. Lower panel: The smoothed Galaxy distribution on a shell at 3750 km s$^{-1}$. Smoothing and contouring are the same as in Figure 5.

**Figure 7**: Upper panel: Galaxy distribution in the 1.2 Jy Survey with recession velocities within 5000 km s$^{-1}$ < $v$ < 7500 km s$^{-1}$. Lower panel: The smoothed Galaxy distribution on a shell at 6250 km s$^{-1}$. Smoothing and contouring are the same as in Figure 5.

**Figure 8**: Upper panel: Galaxy distribution in the 1.2 Jy Survey with recession velocities within 7500 km s$^{-1}$ < $v$ < 10,000 km s$^{-1}$. Lower panel: The smoothed Galaxy distribution on a shell at 8750 km s$^{-1}$. Smoothing and contouring are the same as in Figure 5.

**Figure 9**: The smoothed *IRAS* 1.2 Jy density field out to 10,000 km s$^{-1}$. The center panel is the Supergalactic Plane (SGZ=0). The adjacent panels are parallel to the SG plane and are located beginning at the upper left and proceeding left to right at a) Z = −8000, b) −6000, c) −4000, ..., g) 4000, h) 6000, and i) 8000 km s$^{-1}$. The smoothing kernel is a Gaussian of dispersion varying as the mean interparticle separation. Contour levels are spaced at $\Delta\delta\rho/\rho = 0.25$ with solid (dotted) contours denoted positive (negative) fluctuations about the mean (heavy contour).

**TABLE 1**
OBJECT CLASSIFICATION

|  | Number | |
| --- | --- | --- |
| Source | $f_{60} > 1.2$ Jy | 1.2 Jy $< f_{60} <$ 1.936 Jy |
| Galaxy | 5321 | 2663 |
| Unobserved[a] | 14 | 14 |
| Resolved H II region | 99 | 19 |
| Cirrus | 3057 | 1058 |
| Star | 184 | 58 |
| Emission-line star | 28 | 4 |
| Planetary nebula | 136 | 35 |
| Reflection nebula | 2 | 0 |
| Empty field | 88 | 66 |

Total satisfying selection criteria with $f_{60} >$1.2 Jy: 8934
[a] as of January, 1995

**TABLE 2**
UNOBSERVED OBJECTS[a]

| $\alpha$ (1950) | $\delta$ (1950) | $f_{60}$ (Jy) |
| --- | --- | --- |
| 01:04:41.6 | +39:04:20 | 1.24 |
| 02:33:12.0 | −15:41:43 | 1.60 |
| 04:13:47.1 | +12:17:35 | 1.89 |
| 05:18:34.2 | −13:10:33 | 1.28 |
| 05:51:57.6 | +53:24:55 | 1.26 |
| 06:02:07.4 | −45:09:27 | 1.70 |
| 06:39:38.4 | +44:11:10 | 1.76 |
| 07:31:28.4 | +62:07:45 | 1.44 |
| 07:32:27.2 | +14:23:08 | 1.49 |
| 08:14:37.6 | −60:30:44 | 1.59 |
| 08:53:48.0 | −83:15:44 | 1.22 |
| 10:38:20.6 | +76:37:21 | 1.31 |
| 14:14:12.1 | −70:26:44 | 1.32 |
| 19:12:08.4 | +26:38:50 | 1.30 |

[a] as of January, 1995